\documentstyle[12pt]{article}
\def\tdfullfigure #1 #2 #3 #4 #5 #6
  {\begin{figure}\vspace*{#3 cm}\vspace{12.3cm}
   \tdpsinput #1 #6 #5 #3
   \vspace{-4.0cm}\vspace{#4 cm}
   \caption[]{#2}\end{figure}
   \typeout{TDINPUT: FIGURE \thefigure. produced with file #1}}
\def\tdpsinput#1 #2 #3 #4 {
   \special{ps::
-1 1 scale
-90 rotate
-1700 -2342 translate
#2 #2 scale
#4 -118 mul #3 -118 mul translate
3.5 3.5 scale
      }
   \special{ps: plotfile #1 asis}
   \special{ps:: endexecute
      }}
\begin{document}

\vspace*{2.5cm}

\begin{center}
{\Large {\sc  Dynamical Symmetry Approach to Periodic
    Hamiltonians}}

\vspace{1.8cm}

Hui LI\footnote{
email: huili@nst4.physics.yale.edu} {\rm and} Dimitri KUSNEZOV\footnote{
email: dimitri@nst4.physics.yale.edu}\\

\vspace{1cm}

{\sl Center for Theoretical Physics, Sloane Physics Laboratory,\\Yale
University, New Haven, CT 06520-8120}

\vskip 1.2 cm

{\it April 1999}

\vspace{1.2cm}

\parbox{13.0cm}
{\begin{center}\large\sc ABSTRACT \end{center}
{\hspace*{0.3cm}
We show that dynamical symmetry methods can be applied to
Hamiltonians with periodic potentials. We construct dynamical
symmetry Hamiltonians for the Scarf
potential and its extensions using representations of $su(1,1)$
and $so(2,2)$. Energy bands and gaps are readily understood 
in terms of representation theory. We compute the transfer 
matrices and dispersion relations for these systems, and find
that the complementary series plays a central role as well as
non-unitary representations.}}
\end{center}

\vspace{3mm}

\noindent PACs numbers: 03.65.Fd, 02.20.-a, 02.20.Sv, 11.30.-j\\

\noindent keywords: representation theory, dynamical symmetry, exactly
solvable models, periodic potentials, band structure.

\newpage

\setcounter{page}{2} 

\begin{center}
{\large I. Introduction}
\end{center}
\vspace{7mm}

Lie-algebraic techniques have found wide application to
physical systems and generally provide descriptions of bound
states or scattering states\cite{1,2,AGI}. Once an algebraic
structure is identified, such as a spectrum generating algebra, 
exactly solvable limits of the theory, or dynamical symmetries,
can be constructed\cite{fi1}. Here representation
theory provides a full classification of states and often
transitions\cite{dk}. These dynamical symmetry limits can be 
intuitive guides to the more general structure and behavior of
solutions of the problem. Quantum systems can be
characterized by three types of spectra: discrete (bound
states), continuous (scattering states) and bands (periodic
potentials). The third case corresponds to spectra with 
energy bands and gaps.
Up to now, however, dynamical symmetry treatments have
focused only on the first two, leaving the case of band
structure and its connection to representation theory unclear.

In this article, we extend the dynamical symmetry 
approach to quantum systems by showing that Lie
algebras and representation theory can also be used to treat Hamiltonians
with periodic potentials, allowing the calculation of dispersion
relations and transfer matrices\cite{li}.
We will focus our attention here on the Scarf
potential\cite{scarf} and its generalizations and
show how representations of $so(2,1)$ and
$so(2,2)$ can be used to explain energy bands and gaps. 
The representations which will be necessary are the projective
representations of 
 $su(1,1)\sim so(2,1)$. These have three
families, known as the discrete, principal, and complementary
series. The discrete and principal series have found
much application in physics. For instance, the P\"oschl-Teller
Hamiltonian, $H=-d/dx^2 + g/\cosh^2 x$, can be expressed as an
$su(1,1)$ dynamical symmetry\cite{fi2}, with the discrete
and principal series describing the bound and scattering states.
The complementary series, however, with $-1/2<j<0$, has found
little application in physics and is considered 
 to be more of a curiosity. We
will see that this series is precisely what is needed to
describe band structure in certain periodic potentials, and
further, that the unitary representations correspond to the
energy gaps, rather than the bands.

\vspace{1cm}
\begin{center}
{\large II. Scarf Potential}
\end{center}
\vspace{7mm}

The Scarf potential\cite{scarf} provides a convenient starting
point for the dynamical symmetry analysis of periodic
systems. It was originally introduced as an example of an
exactly solvable crystal model. The starting point is the
Hamiltonian
\begin{equation}
 H_{sc} = -\frac{d^2}{dx^2} + \frac{g}{\sin^2 x}.
\end{equation}
The potential is shown in Fig. 1. (We choose units with mass
$M=1/2$ and $\hbar=1$.)
The strength of the potential $g$ is usually expressed as
$g=s^2-1/4$ since for $g \leq -1/4$, one can no longer define a
Hilbert space for which the Hamiltonian is
self-adjoint\cite{gesztesy1}. The dispersion relation for this
Hamiltonian was found to be
%
\begin{equation}
 E(k) = \frac{1}{\pi^2}\left[ \cos^{-1} \left( \sin\pi s\cos
 k\pi\right)\right]^2
\end{equation}
with the band edges for the $n-$th band:
\begin{equation}
E_n^\pm= (n+\frac{1}{2} \pm s)^2.
\end{equation}
The bands become degenerate as $s\rightarrow 0$. For $s=1/2$, the motion is
that of a free particle with $E(k)=k^2$. While Scarf originally showed
that the potential admits band structure for $0< s \leq 1/2$, it
was demonstrated more recently that the Hamiltonian has
bands for $1/2\leq s<1$\cite{gesztesy1}. In our
analysis, we will see that the entire range  of $0< s < 1$
arises naturally from representation theory.

In order to realize the Scarf problem as a dynamical symmetry,
we consider the Lie algebras isomorphic to $so(3)$. We will see
that while different constructions are possible, not all are fruitful.

\vspace{7mm}
\noindent{\sc A. $so(3)$ realization}
\vspace{7mm}

The relationship of the Scarf Hamiltonian to $so(3)$ was noted
some time ago by G\"ursey\cite{gursey}. Consider the 
realization of $so(3)$ given by the generators:
\begin{eqnarray}
I_{\pm} &=& e^{\pm i \phi} [ \pm\frac{\partial}{\partial \theta} +
\cot\theta (\mp\frac{1}{2} + i \frac{\partial}{\partial \phi})] \\
I_{3} &=& - i \frac{\partial}{\partial \phi} \\
I^{2} &=& I_+I_- + I_3^2 - I_3  \nonumber \\
&=& - \frac{\partial^{2}}{\partial \theta^{2}} -
\frac{1}{\sin^2\theta} 
(\frac{\partial^{2}}{\partial \phi^{2}} + \frac{1}{4}) - \frac{1}{4}
\end{eqnarray}
which satisfy the usual  commutation relations:
\begin{equation}
\lbrack I_{3},I_{+}]=I_{+},\,\ [I_{3},I_{-}]=-I_{-},\,\ [I_{+},I_{-}]=2I_{3}.
\end{equation}
Then, using the basis $\psi _{j}^{m}=\sqrt{\sin \theta }\,P_{j}^{m}(\cos
\theta )$, with the unitary representations of $so(3)$ labeled by
$(j,m)$, the Casimir invariant $I^2$ can be rewritten as the
Schr\"odinger equation:

\begin{equation}
\lbrack -\frac{d^{2}}{d \theta ^{2}}+\frac{m^{2}-\frac{1}{4}}{%
\sin ^{2}\theta }]\,\psi _{j}^{m}(\theta )=(j+\frac{1}{2})^{2}\,\psi
_{j}^{m}(\theta ).
\end{equation}
While this is Scarf's Hamiltonian with $g=m^2-1/2$
(similar to $g=s^2-1/2$ in (1)), it is not a useful
realization for several reasons.
For instance, one cannot
obtain any band structure from the discrete representations of
$so(3)$. Here the spectrum is labeled by $(j+1/2)$, which
identifies only bound states. Further, the strength of the
potential, $m^2-1/4$, is only negative for $m=0$. In this case
$g=-1/4$ and the Hamiltonian is no longer self-adjoint. Finally,
since $m$ appears in the strength $g$ of the potential, a given
representation $j$ would correspond to different forms of the
Hamiltonian, rather than the spectrum of a single Hamiltonian. For this
reason, the previous realizations of $H_{Sc}$ are not useful for
the discussion of band structure.

\vspace{7mm}
\noindent{\sc B. $so(2,1)$ realization}
\vspace{7mm}

A more suitable realization of the Scarf Hamiltonian can be
found using $so(2,1)\sim su(1,1)$. To obtain this form, we
perform the following transformations of the $so(3)$ algebra:
{\it (i)} scaling the wavefunction by $\frac{1}{
\sqrt{\sin \theta }}$ , {\it (ii)} changing $\cos \theta \rightarrow $ $\tanh
\theta $ and {\it (iii)} taking $\theta\rightarrow
i\theta$. The result is the $so(2,1)$ realization
\begin{eqnarray}
I_{\pm } &=&e^{\pm i\phi }(\mp \sin \theta \frac{\partial }{\partial \theta }
+i\cos \theta \frac{\partial }{\partial \phi }) \\
I_{3} &=&-i\frac{\partial }{\partial \phi } \\
I^{2} &=&-I_{+}I_{-}+I_{3}^{2}-I_{3}  \nonumber \\
&=&\sin ^{2}\theta (\frac{\partial ^{2}}{\partial \theta ^{2}}-\frac{%
\partial ^{2}}{\partial \phi ^{2}})
\end{eqnarray}
which satisfies the commutation relations
\begin{equation}
\lbrack I_{3},I_{+}]=I_{+},\,\ [I_{3},I_{-}]=-I_{-},\,\
[I_{+},I_{-}]=-2I_{3}.
\end{equation}
The Casimir operator, using the basis states  
$\psi _{j}^{m}(\theta )=P_{j}^{m}(i\cot \theta ),0<\theta
<\frac{\pi }{2}$, reduces to Scarf's Hamiltonian in the
dynamical symmetry form:

\begin{equation} \label{eq:sds}
\lbrack -\frac{d^{2}}{d\theta ^{2}}+\frac{j(j+1)}{\sin
^{2}\theta }]\,\psi _{j}^{m}(\theta )=m^{2}\,\psi
_{j}^{m}(\theta ) .
\end{equation}

While this Hamiltonian is more pleasing than Eq. (8) in the sense that a
single representation $j$ will account for the spectral
properties, given by $m^2$, the standard unitary representations
(given in Appendix A) are not yet sufficient to describe the bands. 
These come in three series. The principal series with
$j=-1/2+i\rho$, $\rho>0$, the discrete series $D_j^\pm$ where
$j=-n/2$ for $n=1,2,...$, and the complementary series,
$-1/2<j<0$. 

In order to realize band structure as a dynamical symmetry, it
is clear that we must consider slightly more general representations. For
Hamiltonians with periodic potentials, $V(x+a)=V(x)$, Bloch's
theorem requires the form of the wavefunctions to be\cite{kittel}
\begin{equation}\label{eq:bloch}
  \Psi_k(x) = e^{ikx} u_k(x),\qquad u_k(x+a)=u_k(x),
\end{equation}
so that $\Psi_k(x+a)=\exp(ika)\Psi_k(x)$ is not single
valued. To obtain multi-valued functions, we pass
to the projective unitary representations of $su(1,1)\sim
so(2,1)$\cite{bargmann,pukanszky}.
In contrast to the more familiar representations of $so(3)$
which are related to the orthogonal symmetries in the vector
space ${\cal R}^3$, the
projective representations are associated with equivalence classes of
vectors defined up to a phase ( as in Eq. \ref{eq:bloch}). The
action of a group on the projective space (rather than a vector
space), defined by this
equivalence class of states, leads to the projective representations.
While these are multi-valued representations of the group, they
are proper representations of the algebra and are hence
suitable. Consequently, 
the single-valued representations of this covering group of
$su(1,1)$ are infinitely many-valued representations of
$su(1,1)$. Such representations have been used to describe bound
and scattering states in the P\"oschl-Teller
potential\cite{fi2}.
They fall into the same three series as the usual
unitary representations of $su(1,1)$ discussed above (see 
Appendix A). We will see that for our Scarf dynamical symmetry (\ref{eq:sds}),
the discrete series corresponds to the band edges, the
complementary series provides the bands and gaps, while the
principal series is unphysical, corresponding to the regime
where the Hamiltonian is not self-adjoint.

Consider first the complementary series of the projective unitary
representations of $so(2,1)$. Here we must have 
\begin{equation}
  -\frac{1}{2}<j<0,\qquad {\rm or}\qquad -\frac{1}{4}<j(j+1)<0.
\end{equation}
This is precisely the range of $g=j(j+1)$ studied initially by
Scarf in Eq. (1). The states are labeled by two quantum numbers
$j,m$, with unitary representations given by the range of
quantum numbers:

\begin{equation}
  m=m_{0}\pm n\,(n=0,1,\cdots ),
    \qquad 0\leq m_{0}<1,\qquad m_{0}(1-m_{0})<-j(j+1)<\frac{1}{4}.
\end{equation}
The last condition provides the range:
\begin{equation}
  0<m_{0}<-j,\qquad {\rm and}\qquad 1+j<m_{0}<1,
\end{equation}
which is illustrated in Fig. 2. For a given value of $j$,
$j(j+1)$ (dots) separates unitary from non-unitary
representations. The unitary representations are given by values
of $m$ for which the periodically continued parabola (dashes and
solid) are above $j(j+1)$. One can now see that these unitary
representations correspond to the band gaps rather than the
bands by taking $j\rightarrow 0$. In this case the Hamiltonian
(13) is that of a free particle, so that the spectrum is
$E=m^2\ge 0$. From Eqs. (16)-(17) and Fig. 2, we see that as 
$j\rightarrow 0$ the allowed values of $m$  become restricted 
to $m=0,\pm1,\pm2,...$. Therefore, for a specific 
$j$, $E=m^{2}$ has band structure, with the range of $m$ from
unitary projective representations giving the energy gaps.
The {\sl non-unitary} projective representations of the
complementary series give the energy bands
\begin{equation}
  (-j+n)^{2}<E<(1+j+n)^{2},\qquad  n-j<m<1+j+n.
\end{equation}

The band edges are not contained in the complementary
series. In contrast to the states in the band, the edge states
are periodic. They form a discrete set of states which are
associated with the discrete series. These series $D_{j}^{\pm }$ 
have the representations $j<0$ with $m$ given by
\begin{eqnarray}
       D_j^+ \;& : & m=-j,1-j,2-j,...\\
       D_j^- \;& : & m=j,j-1,j-2,...                        
\end{eqnarray}
When we restrict to the range of physical interest,
$-\frac{1}{2}<j<0$, this series provides the upper and lower
band edges (compare to Eq. (18)):
\begin{eqnarray}
   D_j^\pm ({\rm lower}) &: & E=(n-j)^2\\
   D_{-j-1}^\pm ({\rm upper}) &: & E=(n+j+1)^2.
\end{eqnarray}
Eq. (22) arises from the invariance of our Hamiltonian (13)
under $j\rightarrow -1-j$, allowing both discrete series
$D^\pm_j$ and $D^\pm_{-1-j}$.
Other discrete representations with $j<-1$ are not useful for band
structure. The band spectrum of the Scarf potential which
includes both the discrete and complementary series is shown in
Fig. 3. The shaded region corresponds to the bands (non-unitary) and the
unshaded to the gaps (unitary).
 
The remaining representations, the principal series, has
$j=-\frac{1}{2}+i\rho$  $(\rho >0)$. This gives a potential with strength
$g=j(j+1)<-\frac{1}{4}$, for which the Hamiltonian is no longer
self-adjoint and is of no physical interest.

Note that we have explained the band structure for strengths of
the potential $-1/4 < g=j(j+1)<0$ and found agreement with
Scarf\cite{scarf}. More recently it was noted that for
$0 \le g < 3/4$, there is also band
structure \cite{gesztesy1}.  In this range the potential is
strictly positive. (The origin of the band structure here is that
the matching conditions on the wavefunctions around the
singularity in the potential, needed to have a self-adjoint
Hamiltonian, in a sense `dilute' the infinite potential at these
points and allow bands.) While our $so(2,1)$ realization above cannot
account for this range of $g$, we will see in Section III, that a
limiting case of an $so(2,2)$ dynamical symmetry will account
for this range using the same complementary series.  For 
$g \ge 3/4$, there is no band structure and the discrete projective
representation then describe the bound state spectrum.

\vspace{7mm}
\noindent{\sc C. Transfer matrix}
\vspace{7mm}

The transfer matrix T for the period $x\in (-\frac{\pi }{2},\frac{\pi
}{2})$ can be computed directly from wave functions. However,
the quadratic singularity of the potential
requires some care. There are two approaches one can consider, but both are
equivalent\cite{scarf,gesztesy1,james}. In the first, we compute
the transfer matrix at $x=\pm\varepsilon$. We then match the
transfer matrices on both sides of the singularity as
$\varepsilon\rightarrow 0$, which results in matching conditions
on the wavefunctions. This procedure is not equivalent to an analytical
continuation around the origin. The second arises in the
construction of the
Hilbert space of functions for which $H$ is self-adjoint. This
gives rise to equivalent matching conditions around the origin\cite{gesztesy1}.
The matrix elements of the transfer matrix are related to the values of the
even and odd solutions and their first derivatives at 
$\frac{\pi}{2}$ (see Appendix B). We find

\begin{equation}
T \, = \, \left(
\begin{array}{cc}
\alpha & \beta \\
\beta^{*} & \alpha^{*}
\end{array}
\right)
\end{equation}
where $\alpha$ and $\beta$ are determined by the representations
of the complementary and discrete series $j,m$ as:

\begin{eqnarray}
\alpha &=& e^{- i m \pi} \, \left[ \frac{\cos\pi m}{\sin\pi(j+\frac{1}{2})}
               \right.\nonumber \\
      & &        \left. + \, i \left(-\frac{2}{m} \frac{ 
           \Gamma (\frac{1-j+m}{2}) \Gamma (\frac{1 -j-m}{2} )}{
           \Gamma (- \frac{j-m}{2}) \Gamma ( - \frac{j+m}{2} )} + \frac{m}{2}
           \frac{ \Gamma ( \frac{1+j+m}{2}) 
           \Gamma ( \frac {1+j - m}{2})}{ \Gamma (\frac{2+j+m}{2} ) 
          \Gamma ( \frac{2 +j -m}{2})}\right)\frac{\cos\pi
               \frac{j+m}{2} \; \cos\pi
             \frac{j-m}{2}}{\sin\pi(j+\frac{1}{2})} \right]\\
\beta &=& i \, e^{i m \pi} \, \frac{\cos\pi \frac{j+m}{2} \/
               \cos\pi \frac{j-m}{2}}{\sin\pi(j+\frac{1}{2})}  \nonumber \\
& &  \left[\frac{2}{m} \frac{ \Gamma
               (\frac{1-j+m}{2})\Gamma(\frac{1-j-m}{2})}
               {\Gamma ( -\frac{j-m}{2}) \Gamma ( -
               \frac{j+m}{2})} + \frac{m}{2} \frac{\Gamma (
               \frac{1+j+m}{2}) 
              \Gamma (\frac{1+j-m}{2})}{ \Gamma (
               \frac{2+j+m}{2}) \Gamma ( \frac{2+j-m}{2})}\right] .
\end{eqnarray}
Although the Scarf Hamiltonian can be obtained from the
P\"oschl-Teller potential $V(x)=g/\cosh^2x$ through a
transformation, the above transfer matrix is not related to that
of the P\"oschl-Teller in any simple manner.

The Bloch form of the $so(2,1)$ wave functions for the $n-$th
period, $(n-\frac{1}{2})\pi < x \leq (n+\frac{1}{2})\pi$, 
of the Scarf Hamiltonian are readily found to be
\begin{equation}
\Psi_k (x) = f_k (x-n\pi) e^{ikx}
\end{equation}
where:
\begin{equation}
f_k(z) = e^{-ik(z+\frac{\pi}{2})}[a P_j^m (i\cot z) +
      b P_j^{-m}(i\cot z)]
\end{equation}
and:
\begin{eqnarray}
a &=& (-)^{-j/2}\frac{\sqrt{\pi}2^{-m}}{\sin m\pi} \left[\cos\frac{k\pi}{2}
\frac{\Gamma(\frac{1-j-m}{2})}{\Gamma(\frac{-j+m}{2})} 
 - \sin\frac{k\pi}{2}
\frac{\Gamma(\frac{2+j-m}{2})}{\Gamma(\frac{1+j+m}{2})} (-)^j\right] \\
b &=& -(-)^{-j/2}\frac{\sqrt{\pi}2^m}{\sin m\pi} \left[\cos\frac{k\pi}{2}
\frac{\Gamma(\frac{1-j+m}{2})}{\Gamma(\frac{-j-m}{2})} 
  - \sin\frac{k\pi}{2}
\frac{\Gamma(\frac{2+j+m}{2})}{\Gamma(\frac{1+j-m}{2})} (-)^j \right].
\end{eqnarray}
Since $-\frac{\pi}{2} < z\leq \frac{\pi}{2}$, $f_k(z)$ is made
periodic, and $\Psi_k(x)$ satisfies Bloch's theorem.

\vspace{7mm}
\noindent{\sc  D. Dispersion relation}
\vspace{7mm}

Once we have the transfer matrix,  the dispersion relation is 
obtained from $\alpha$ by the condition\cite{kittel,cohen}:
\begin{equation}
\cos \pi k \, = \, Re(\alpha e^{im\pi}) \, = \, \frac{\cos \pi m}{\sin \pi
(j + \frac{1}{2})} \, .
\end{equation}
Solving for the energy $E=m^2$, we find:
\begin{equation}
E(k) = m^2 = \frac{1}{\pi^2}[\cos^{-1}(\sin\pi(j+\frac{1}{2})\cos\pi k)]^2 .
\end{equation}
This is precisely the result (2) obtained by Scarf. Again, the
values of $j$ and $m$ are determined from the representations
given in (18) and (21)-(22). From the dispersion
relation, we can also compute the group velocity ${\cal V}$ and
the effective mass $M^*$. These will depend only upon the
representation labels $j$ and $m$. We have:
\begin{equation}
{\cal V}(j,m) = \frac{\partial E}{\partial
  k} = 2 m \frac{\sqrt{\cos^2 \pi j -\cos^2\pi m}}{\sin \pi m}
\end{equation}
This is plotted in Fig. 4(a) for selected values of $j$. ${\cal
  V}(j,m)$ vanishes on the band edges. For
$j=0$, the Hamiltonian (13) describes free motion, and we expect
${\cal   V}=\pm k/M=\pm 2k$ (dots), while for $j\rightarrow-1/2$, we
  have degenerate bands,
and ${\cal V}\rightarrow 0$ at half-integer values of $m$. 
For the effective mass:
\begin{eqnarray}
\frac{1}{M^*(j,m)} &=& \frac{\partial^2 E}{\partial
 k^2}\nonumber \\
 &=& 2\left[\frac{\cos^2j\pi}{\sin^2 m\pi} - \cot^2 m\pi +
 m\pi\frac{\sin^2j\pi \cot m\pi}{\sin^2 m\pi}\right].
\end{eqnarray}
(Note that this differs slightly from the result derived in \cite{scarf}.)
In Fig. 4(b),  $1/M^*(j,m)$ is shown for selected values of $j$. For
$j= 0$,  $M^*= M=1/2$, while for
$j\rightarrow -1/2$, $1/M^*\rightarrow 0$.

\vspace{7mm}
\noindent{\sc E. Variation of Scarf potential}
\vspace{7mm}

In the next section we will present a dynamical symmetry
Hamiltonian for a variation of the Scarf potential using
$so(2,2)$. This potential will have several limits where the
Hamiltonian reduces to the Scarf case, including the
$1/cos^2x$ potential. In order to compare the transfer matrix in this
limit to the Scarf result, we consider the Scarf Hamiltonian
translated by $\frac{\pi }{2}$,
\begin{equation}
\lbrack -\frac{d ^{2}}{d \theta ^{2}}+\frac{j(j+1)}{\cos
^{2}\theta }]\,\psi _{j}^{m}(\theta )=m^{2}\,\psi
_{j}^{m}(\theta ).
\end{equation}
The dispersion relation $E(k)$ and the energy band structure
will remain the same as before. The
transfer matrix for $(-\frac{\pi }{2},\frac{\pi }{2})$, on the
other hand, will change. The new transfer matrix 
can be calculated easily from a translation of the solutions of Scarf case:

\begin{eqnarray}
\alpha &=&e^{-im\pi }\, \left[\frac{\cos \pi m}{\sin \pi (j+\frac{1}{2})}
+\,i\left(-m \frac{\Gamma (j+\frac{1}{2})\Gamma (\frac{1-j+m}{2})\Gamma
(\frac{ 1-j-m}{2})}{\Gamma (-j-\frac{1}{2})\Gamma
(\frac{2+j+m}{2})\Gamma (\frac{ 2+j-m}{2})}\right.\right.
\nonumber \\
 & & \left.\left.+\frac{1}{m}\frac{\Gamma (-j-\frac{1}{2})\Gamma
(\frac{1+j+m}{2} )\Gamma (\frac{1+j-m}{2})}{\Gamma
(j+\frac{1}{2})\Gamma (-\frac{j-m}{2} )\Gamma (-\frac{j+m}{2})}\right)
\frac{\cos \pi \frac{j+m}{2}\/\cos \pi
\frac{j-m}{2}}{\sin \pi (j+\frac{1}{ 2})}\right]\\
\beta &=& -i\,e^{im\pi }\,\frac{\cos \pi \frac{j+m}{2}\/\cos \pi \frac{j-m}{2}
}{\sin \pi (j+\frac{1}{2})} \left[m\frac{\Gamma (j+\frac{1}{2})\Gamma
(\frac{1-j+m}{2})\Gamma (\frac{1-j-m}{ 2})}{\Gamma
(-j-\frac{1}{2})\Gamma (\frac{2+j+m}{2})\Gamma (\frac{2+j-m}{2})}\right.
\nonumber \\
 && \left. +\frac{1}{m}\frac{\Gamma (-j-\frac{1}{2})\Gamma
(\frac{1+j+m}{2})\Gamma (\frac{1+j-m}{2})}{\Gamma
(j+\frac{1}{2})\Gamma (-\frac{j-m}{2})\Gamma (-
\frac{j+m}{2})}\right]
\end{eqnarray}

\vspace{1cm}
\begin{center}
{\large III. Generalized Scarf Potential}
\end{center}
\vspace{7mm}

We have now shown that band structure can arise naturally as a
dynamical symmetry. We would like to build on the analysis of
the Scarf problem and study a different class of
periodic potentials. Consider an extension of the Scarf potential
given by a generalized P\"oschl-Teller Hamiltonian\cite{poschl}
\begin{equation}
[-\frac{d^{2}}{d x^{2}} + \frac{g_1}{\sin^2 x} +
\frac{g_2}{\cos^2 x} ] \, \Psi(x) \, = \, E \, \Psi(x),
\qquad (g_1,g_2 > -\frac{1}{4})
\end{equation}
While this Hamiltonian is exactly solvable, we would like to see
how band structure can be obtained from representation theory
using dynamical symmetry considerations.
We will relate this Hamiltonian to the $so(4)$ and
$so(2,2)$  algebras and develop the band structure from the
complementary series. We plot some forms of this potential in
Fig. 5 for several values of $g_1$ and $g_2$. Our study will be
restricted to the range $-1/4< g_1,g_2\leq 0$. 

\vspace{7mm}
\noindent{\sc A. $so(4)$ realization}
\vspace{7mm}

We start with the realization of the $so(4)$ algebra:

\begin{eqnarray}
A_{\pm } &=&\frac{1}{2}e^{\pm i(\phi +\alpha )} \left[\pm
     \frac{\partial }{\partial \theta }+\cot 2\theta
    \left(i\frac{\partial }{\partial \phi }+ i\frac{\partial
     }{\partial \alpha }
          \mp 1\right) -\frac{i}{\sin
    2\theta }\left(\frac{\partial }{\partial \phi }-\frac{\partial
     }{\partial \alpha }\right)\right]  \nonumber \\
A_{3} &=&-\frac{i}{2}\left(\frac{\partial }{\partial \phi }+\frac{\partial }{%
       \partial \alpha }\right) \\
B_{\pm } &=&\frac{1}{2}e^{\pm i(\phi -\alpha )}
\left[\pm\frac{\partial}{\partial\theta}
   +\cot 2\theta \left(i\frac{\partial }{\partial \phi
       }-i\frac{\partial }{\partial \alpha }\mp
     1\right)
    -\frac{i}{\sin2\theta } \left(\frac{\partial }{\partial \phi }+\frac{\partial
     }{\partial \alpha }\right) \right]\nonumber\\
B_{3} &=&-\frac{i}{2}\left(\frac{\partial }{\partial \phi
    }-\frac{\partial }{\partial \alpha }\right)\nonumber
\end{eqnarray}
which have the commutation relations:
\begin{equation}
\begin{array}{cccc}
{\lbrack A_{3},A_{+}]}=A_{+},& {[A_{3},A_{-}]}=-A_{-},
&{[A_{+},A_{-}]}=2A_{3}, &\\
{\lbrack B_{3},B_{+}]}=B_{+},& {[B_{3},B_{-}]}=-B_{-}, &
{[B_{+},B_{-}]}=2B_{3}, & {[A,B]}=0.\end{array}
\end{equation}
Since this is the direct product of two $so(3)$ algebras,  the quadratic Casimir
invariant has the form:

\begin{eqnarray}
C_{2} &=&2(A^{2}+B^{2})  \nonumber \\
&=&2(A_{+}A_{-}+A_{3}^{2}-A_{3}+B_{+}B_{-}+B_{3}^{2}-B_{3}) \\
&=&-\frac{\partial ^{2}}{\partial \theta ^{2}}+\frac{1}{\cos ^{2}\theta }\left[
-\frac{\partial ^{2}}{\partial \phi ^{2}}-\frac{1}{4}\right]
+\frac{1}{\sin ^{2}\theta }\left[-\frac{\partial ^{2}
}{\partial \alpha ^{2}}-\frac{1}{4}\right]-1 \nonumber
\end{eqnarray}

The representations of $so(4)$ can be labeled by
($j_1,m$; $j_2,c$), where $j_1,j_2,m,c$ are non-negative
integers or half
integers and $-j_1 \le m \le j_1$, $-j_2 \le c \le j_2$. It is easy
to check that, as differential operators, $A^2=B^2$. So for this
realization, we only need to consider symmetric representations
with $j_1=j_2=j$. Hence,  $C_2=4j(j+1)$. The resulting Schr\"odinger
equation is

\begin{equation}
\lbrack -\frac{d^{2}}{d \theta ^{2}}+\frac{(m+c)^{2}-\frac{1}{%
4}}{\cos ^{2}\theta }+\frac{(m-c)^{2}-\frac{1}{4}}{\sin ^{2}\theta }]\,\psi
_{j}^{m,c}(\theta )=(2j+1)^{2}\,\psi _{j}^{m,c}(\theta )
\end{equation}
While this is suitable for bound states,  the discrete representations of
$so(4)$ do not explain band structure, and the strength of the
potential is not in the range of physical interest.

\vspace{7mm}
\noindent{\sc B. $so(2,2)$ realization}
\vspace{7mm}

We can derive a more suitable realization by passing to
$so(2,2)$. Starting with the above generators, we {\it (i)}
scale the wavefunctions by
$\frac{1}{\sqrt{\sin\theta}}$ , {\it (ii)} transform 
$\cos\theta \rightarrow \tanh \theta$ and {\it (iii)} take
$\theta\rightarrow i\theta$. This results in the $so(2,2)$ realization:

\begin{eqnarray}
A_{\pm} &=& \frac{1}{2} e^{\pm i (\phi + \alpha)} \, \left[ \pm \cos \theta \frac{
         \partial}{\partial \theta}  + \, i \frac{1 +
          \sin^2\theta}{2\sin\theta}
         \left(\frac{\partial}{\partial\phi} 
          + \frac{\partial}{\partial \alpha}\right)\right. \nonumber \\
& & \left.+ i \frac   {\cos^2\theta}{2\sin\theta} 
       \left(\frac{\partial}{\partial \phi} - \frac{\partial
        }{\partial \alpha}\right) \mp \frac{1}{2\sin\theta} \right] \nonumber\\
A_{3} &=& - \frac{i}{2} \left(\frac{\partial}{\partial \phi} + \frac{\partial}{
                \partial \alpha}\right) \\
B_{\pm} &=& \frac{1}{2} e^{\pm i (\phi - \alpha)} \, \left[ \pm 
              \cos\theta\frac{
               \partial}{\partial \theta}  + \, i \frac{1 +
               \sin^2\theta}{2\sin\theta} \left(\frac{\partial}{\partial\phi} 
              -\frac{\partial}{\partial \alpha}\right)\right. \nonumber \\
& & \left.+ i \frac
{\cos^2\theta}{2\sin\theta} \left(\frac{\partial}{\partial \phi} + \frac{\partial
}{\partial \alpha}\right) \mp \frac{1}{2\sin\theta} \right] \nonumber\\
B_{3} &=& - \frac{i}{2} \left(\frac{\partial}{\partial \phi} - \frac{\partial}{
\partial \alpha}\right)\nonumber
\end{eqnarray}
with the commutation relations
\begin{equation}
\begin{array}{cccc}
{\lbrack A_{3},A_{+}]}=A_{+},& {[A_{3},A_{-}]}=-A_{-},
&{[A_{+},A_{-}]}=-2A_{3}, &\\
{\lbrack B_{3},B_{+}]}=B_{+},& {[B_{3},B_{-}]}=-B_{-}, &
{[B_{+},B_{-}]}=-2B_{3}, & {[A,B]}=0.\end{array}
\end{equation}
The quadratic Casimir invariant now has the form
\begin{eqnarray}
C_{2} &=&2(A^{2}+B^{2})  \nonumber \\
&=&2(-A_{+}A_{-}+A_{3}^{2}-A_{3}-B_{+}B_{-}+B_{3}^{2}-B_{3}) \\
&=&\cos ^{2}\theta \frac{\partial ^{2}}{\partial \theta ^{2}}-\cos
^{2}\theta \frac{\partial ^{2}}{\partial \alpha ^{2}}+\frac{\cos ^{2}\theta
}{\sin ^{2}\theta }(\frac{\partial ^{2}}{\partial \phi ^{2}}+\frac{1}{4})-%
\frac{3}{4} \nonumber 
\end{eqnarray}

The states of the representations of $so(2,2)$ can be labeled by
a direct product of representations of $so(2,1)$, denoted 
($j_1,m$; $j_2,c$). Again, as
differential operators, $ A^2 = B^2$ so that $j_1 = j_2 = j$.
Replacing $\theta$ by $x$, this leads to the Schr\"odinger equation:

\begin{equation}
\lbrack -\frac{d^{2}}{d x^{2}}+\frac{(m+c)^{2}-\frac{1}{%
4}}{\sin ^{2}x }+\frac{(2j+1)^{2}-\frac{1}{4}}{\cos ^{2}x }]\,\psi
_{j}^{m,c}(x )=(m-c)^{2}\,\psi _{j}^{m,c}(x )
\end{equation}
Two independent solutions \cite{gesztesy2,miller} in the
region  $0< x <\frac{\pi}{2}$ are:

\begin{eqnarray}
\psi_1(x) &=& (\sin^2 x)^{\frac{1}{4}-\frac{m+c}{2}}
(\cos^2 x)^{-j-\frac{1}{4}} {}_2F_1(-c-j,-m-j;1-m-c; \sin^2 x) \\
\psi_2(x) &=& (\sin^2 x)^{\frac{1}{4}+\frac{m+c}{2}}
  (\cos^2 x)^{-j-\frac{1}{4}} {}_2F_1(m-j,c-j;1+m+c; \sin^2 x) \nonumber 
\end{eqnarray}

In order to develop the band structure of this Schr\"odinger
equation, we must construct the complementary series of the
projective representations of $so(2,2)\sim su(1,1)\oplus
su(1,1)$. This direct product structure allows us to simply use
the results discussed in the Scarf dynamical symmetry.

The complementary series, labeled by $(j,m,c)$, is constructed as
follows. For ranges of $m$ and $c$ which correspond to unitary
representations of (projective) complementary series $su(1,1)$, 
the resulting $so(2,2)$ representation is also
unitary. For ranges of $m$ and $c$ which are both non-unitary,
the resulting direct product becomes unitary in the strip
of physical interest, $0<|m+c|\leq 1/2$ . The
remaining cases when $m$ is unitary and $c$ is non-unitary and
the case with $m$ and $c$ interchanged, result in non-unitary
representations of the complementary series of $so(2,2)$. These
non-unitary representations correspond to the energy bands of the extended
Scarf potential, which can be seen by taking limiting cases
where (i) the potential reduces to the Scarf case (see below)
and (ii) the potential vanishes and the spectrum is continuous.

Since the eigenvalue of our Hamiltonian is $E=(m-c)^2$, and the
strength of the potential is labeled by $j$ and $m+c$, it is
convenient to plot the resulting unitary and non-unitary
representations of $so(2,2)$ versus  $m+c$ for selected
values of $j$. This is done in Fig. 6. Here the energy gaps
correspond to the shaded regions and the bands to the unshaded
regions. Three values of $j$ are chosen: (a) $j=-0.45$, (b)
$j=-0.35$ and (c) $j=-0.25$. Case (c) corresponds to the Scarf
potential limit. As $j\rightarrow -1/2$ or $|m+c|\rightarrow 0$,
the bands become degenerate. On the other hand, when
$j\rightarrow -1/4$ and
 $|m+c|\rightarrow 1/2$, the spectrum
becomes continuous. For the band edges, one takes the direct
product of $su(1,1)$ discrete projective representations.

The bands $E=(m-c)^2$ are given by the  following ranges of
quantum numbers in the  $(m,c)$ plane:

\begin{eqnarray}
2n -(m_o+c_o)-2j &\leq & m-c\leq 2n+1-\mid 2j+1-m_o-c_o\mid,\\
2n+1+\mid 2j+1-m_o-c_o\mid &\leq & m-c\leq 2n+2+2j+m_o+c_o,\nonumber
\end{eqnarray}
where $n=0,1,2,...$ and
\begin{equation}
  0< \mid m+c\mid \leq \frac{1}{2},\qquad 0< 2j+1\leq \frac{1}{2}.
\end{equation}

\vspace{7mm}
\noindent{\sc C. Transfer matrix}
\vspace{7mm}

Due to the strong singularity structure of the potential, one
again must introduce boundary conditions for the solutions at
singularities such that the Schr\"odinger operator can be made
self-adjoint. Such an analysis has been undertaken in Refs. 
\cite {gesztesy1,gesztesy2}.  We can then easily
compute the transfer matrix for the interval $x\in
(-\frac{\pi}{2}, \frac{\pi}{2})$
using the boundary values and first derivatives at $\frac{\pi}{2}$.
The transfer matrix is:

\begin{equation}
T\,=\,\left(
\begin{array}{cc}
\alpha & \beta \\
\beta ^{\ast } & \alpha ^{\ast }
\end{array}
\right)
\end{equation}
where

\begin{eqnarray}
\alpha &=& e^{- i (m-c) \pi} \, \left[ \frac{\cos\pi (m-c)  + 
                   \cos \pi (2j+1) \cos\pi (m+c)}{\sin\pi(2j+1)
                   \sin\pi(m+c)}\right.  \nonumber \\
& & \: + \, i \left( \frac{1}{m-c} \frac{ \Gamma(-2j-1)\Gamma (1+j-m) \Gamma (1
+j -c)}{ \Gamma (2j+1) \Gamma (-j-m)\Gamma(-j-c)}\right.  \nonumber \\
& & \left.- \, (m-c) \frac{ \Gamma (2j+1 ) \Gamma (-j+m)\Gamma(-j+c)}{
\Gamma(-2j-1)\Gamma (1+j+m) \Gamma (2 +j +c)}\right)  \nonumber \\
& & \left.\frac{\sin\pi(j-m)\sin\pi(j-c)}{\sin\pi(2j+1)\sin\pi(m+c)}\right]\\
\beta &=& -i \; e^{i (m-c) \pi}\frac{\sin\pi(j-m)\sin\pi(j-c)}{%
\sin\pi(2j+1)\sin\pi(m+c)}  \nonumber \\
& & \left[\frac{1}{m-c} \frac{ \Gamma(-2j-1)\Gamma (1+j-m) \Gamma (1 +j -c)}{
\Gamma (2j+1) \Gamma (-j-m)\Gamma(-j-c)} \right. \nonumber \\
& & + \, \left.(m-c) \frac{ \Gamma (2j+1 ) \Gamma (-j+m)\Gamma(-j+c)}{
\Gamma(-2j-1)\Gamma (1+j+m) \Gamma (2 +j +c)}\right]
\end{eqnarray}

\vspace{7mm}
\noindent{\sc D. Dispersion relation}
\vspace{7mm}

The dispersion relation is computed as before, using $\cos \pi k
= Re(\alpha e^{i(m-c)\pi})$:

\begin{equation}
\cos\pi k = \frac{\cos\pi (m-c) + \cos \pi (2j+1) \cos \pi (m+c)}{%
\sin\pi(2j+1)\sin\pi(m+c)} .
\end{equation}
If we denote
\begin{equation}
  m_+ = m+c,\qquad m_-=m-c,
\end{equation}
then $E=(m-c)^2=m_-^2$, and we find:

\begin{eqnarray}
E(k) &=&  m_-^2  \nonumber \\
&=& \frac{1}{\pi^2}\left[\cos^{-1}(\cos\pi k \sin\pi(2j+1)\sin\pi m_+
 -\cos\pi(2j+1)\cos\pi m_+)\right]^2
\end{eqnarray}

The band structure could be explained through the projective representations
of $so(2,2)$ when $0<|m+c|\leq \frac{1}{2}$ and $-\frac{1}{2} < j \leq
-\frac{1}{4}$. Again, non-unitary
representations give the energy bands while unitary representations
correspond to energy gaps.

The group velocity for this potential is 
\begin{eqnarray} 
{\cal V}(j,m,c) &=& \frac{\partial E}{\partial
  k}\nonumber\\
 & =& \frac{2 m_-}{\sin \pi  m_-} \left[\sin^2(2\pi j) - \cos^2
  \pi  m_+ -\cos^2 \pi  m_-\right.\\
 & & \left.+ 2\cos\pi m_-\cos\pi m_+\cos 2\pi j\right]^{1/2}\nonumber
\end{eqnarray}
The behavior is shown in Fig. 7 for selected values of $j$ and
$m+c$ given by the dashed lines in Fig. 6.
The effective mass  $M^*(j,m,c)$ is given by:
\begin{eqnarray}
\frac{1}{M^*} &=& \frac{\partial^2 E}{\partial k^2}\nonumber\\
 & =& 2  m_-\pi \left[ \cot\pi m_- - \cos 2\pi j\cos
              \pi m_+\csc\pi m_-\right]\nonumber \\
 & & +  2\csc^2\pi m_-\left(1- m_-\pi\cot\pi
          m_-\right)\left(\sin^22\pi j - \cos^2\pi m_+ \right.\\
 & & \left. - \cos^2\pi m_- +
        2\cos\pi m_-\cos\pi m_+\cos 2\pi j\right) \nonumber
\end{eqnarray}
%

\vspace{7mm}
\noindent{\sc E. Limiting cases}
\vspace{7mm}

There are three cases where the extended Scarf potential reduces
to the Scarf case: {\it (i)} When $2j+1 = \frac{1}{2}$, the potential
becomes the Scarf potential and the transfer matrix is
equivalent to Eqs. (24)-(25). {\it (ii)} When $m+c=\frac{1}{2}$,
Eq. (38) reduces to the
potential
\begin{equation}
  \frac{ (2j+1)^2 - \frac{1}{4}}{\cos^2\theta}
\end{equation}
and the transfer matrix is consistent with the results of Sec. II.E.
{\it (iii)} When $|m+c|=2j+1$, the Hamiltonian
reduces to the Scarf potential with twice the period. 

Of the three limiting cases, it is case {\it (ii)} which
provides something new. To compare to the Scarf results, we let
$2j+1=\tilde j+\frac{1}{2}$, so that the potential (57) becomes
$\tilde j(\tilde j+1)/\sin^2x$. For the full complementary
series $-1/2< j\leq 0$, we have $-1/2< \tilde j\leq 1/2$ which
corresponds to potentials $g/\sin^2 x$ with $-1/4< g <
3/4$. From (47) we find that the energy bands are given by

\begin{eqnarray}
2n-\tilde j &\leq & m-c\leq 2n+1+\tilde j,\\
2n+1-\tilde j &\leq & m-c\leq 2n+2 + \tilde j.\nonumber
\end{eqnarray}
This is precisely the band structure obtained in Eq. (18),
(21)-(22), but now extended to positive coupling constants,
$0\leq g < 3/4$, while using the same complementary series. This
agrees with the more recent observation that the Scarf potential
admits band structure for ranges of the strength which are 
positive\cite{gesztesy1}. It also exemplifies the fact that a
dynamical symmetry does not necessarily exhaust all
possible regimes of band structure, and that other realizations
might provide additional regions. In principle we can
extend our analysis of the generalized Scarf potential to
$g_1,g_2 >0$ as well, but we do not do so here.

\vspace{1cm}
\begin{center}
{\large IV. Conclusions}
\end{center}
\vspace{7mm}

We have shown that dynamical symmetry techniques can be applied to
Hamiltonians with periodic potentials, and band structure can arise
naturally from representation theory. This fills a long-standing gap in the
algebraic approach to quantum systems. We have constructed
dynamical symmetry Hamiltonians in $so(2,1)$ and $so(2,2)$ which
can be expressed as Schr\"odinger operators with periodic
potentials. Using projective representations motivated by Bloch's theorem,
we have seen that the complementary series of $so(2,1)$ and $so(2,2)$
(and their {\it non-unitary} representations) are needed to 
explain band structure, while the discrete representations are
important for band edges. As far as we know, this is the first
application of the $su(1,1)$ complementary series to a physical
problem. It now seems reasonable to loosely associate the
three series of projective representations, discrete, principal
and complementary, with the quantum problems of bound states,
scattering states and energy bands.

  Using our dynamical symmetries, Hamiltonians such as Scarf's
and its extension can be reduced to quadratic forms of the
Cartan subalgebra generators, such as $H=J_z^2$,
which are readily solved. We are then able to derive not only the band
structure, but the dispersion relation and transfer matrix as
well. It would be interesting to develop
higher dimensional periodic Hamiltonians connected to
representations of $u(n,m)$ or $so(n,m)$. In this case, the
inclusion of additional discrete symmetries using point groups
would be possible, and extensions to non-dynamical symmetry
problems could be pursued.

\vspace{7mm}
\noindent{\bf ACKNOWLEDGMENTS}
\vspace{7mm}

We would like to thank F. Iachello for many useful discussions. This work
was supported by DOE grant DE-FG02-91ER40608.

\newpage

\appendix
\renewcommand{\thesection}{{\bf APPENDIX} \Alph{section}.}
\setcounter{equation}{0}
\setcounter{section}{1}
\renewcommand{\theequation}{\Alph{section}\arabic{equation}}

\begin{center}
{\large Appendix A: Representations of $so(2,1)$}
\end{center}
\vspace{7mm}

First, let us recall the presentation of $so(3)$. The algebra can
be realized as differential operators on the sphere $x^2 + y^2 +
z^2 = 1$. The representations are labeled by $(j,m)$ where $j$ is any
non-negative half integer and $-j \le m \le j$.

The $so(2,1)$ algebra can be realized as differential operators on a
hyperboloid $- x^2 - y^2 + z^2 = 1$. The unitary representations are \cite
{bargmann}:

\begin{itemize}
\item  The principal series $j=-\frac{1}{2}+i\rho ,\,\rho >0,\,m=0,\pm
1,\ldots $ or $m=\pm \frac{1}{2},\pm \frac{3}{2},\ldots $.

\item  The complementary series $-\frac{1}{2}<j<0,\,m=0,\pm 1,\ldots $.

\item  The discrete series $D_{j}^{+}$, where $j$ is a negative integer or
half integer and $m=-j,-j+1,\ldots $.

\item  The discrete series $D_{j}^{-}$, where $j$ is a negative integer or
half integer and $m=j,j-1,\ldots $.
\end{itemize}

A more general form of the representations of the algebra are
 the projective representations. The projective unitary
representations of $so(2,1)$ are \cite{pukanszky}:

\begin{itemize}
\item  The principal series $j=-\frac{1}{2}+i\rho ,\,\rho >0,\,0\leq
m_{0}<1,\,m=m_{0}\pm n,\,n=0,1,2,\ldots $

\item  The complementary series $-\frac{1}{2}<j<0,\,0\leq m_{0}<1,\
,m_{0}(m_{0}-1)>j(j+1)\geq -\frac{1}{4},\,m=m_{0}\pm n,n=0,1,\ldots $.

\item  The discrete series $D_{j}^{+},j<0$, $m=-j, -j+1,...$

\item  The discrete series $D_{j}^{-},j<0$, $m=j, j-1,...$
\end{itemize}

Since we find that the non-unitary representations are important
for the bands, we review their origin\cite{pukanszky}.
Assuming $I_3 f = m_0 f$, with $0\leq m_0< 1$, and using the 
commutation relations for  $ so(2,1) $, we have
\begin{eqnarray}
I_{3}I_{+}f &=&(m_0 +1)I_{+}f \\
I_{3}I_{-}f &=&(m_0 -1)I_{-}f \\
I_{-}I_{+}f &=&[-j(j+1)+m_0 (m_0 +1)]f \\
I_{+}I_{-}f &=&[-j(j+1)+m_0 (m_0 -1)]f
\end{eqnarray}
where $I^{2}=j(j+1)$ is the Casimir, a constant for a specific
representation.
Replacing $f$ by $I_{+}^{n-1}f$ and $I_{-}^{n-1}f\,(n=1,2,\dots )$ in
the last two equations, we get

\begin{eqnarray}
I_{-}I_{+}^{n}f &=&\alpha _{n}I_{+}^{n-1}f \\
I_{+}I_{-}^{n}f &=&\beta _{n}I_{-}^{n-1}f
\end{eqnarray}
where $\alpha _{n}=-j(j+1)+(m_0 +n-1)(m_0 +n)$ and 
$\beta _{n}=-j(j+1)+(m_0-n)(m_0 -n+1)$. The above relations imply
\begin{eqnarray}
||I_{+}^{n+1}f||^{2} &=&(I_{+}^{n+1}f,I_{+}^{n+1}f)=\alpha
_{n+1}||I_{+}^{n}f||^{2} \\
||I_{-}^{n+1}f||^{2} &=&(I_{-}^{n+1}f,I_{-}^{n+1}f)=\beta
_{n+1}||I_{-}^{n}f||^{2}
\end{eqnarray}

Starting with the initial state $f$, we can generate the
coefficients $\alpha_k$ and $\beta_k$ ($k>0$). Of these
coefficients, only $\beta_1$ can be positive or negative.
This distinguishes the unitary and non-unitary
representations. For instance $\beta_1>0$ when
$m_0(m_0-1)>j(j+1)$, which gives the complementary series. When
we are in the region $-j< m_0< 1+j$, $\beta_1< 0$. So
if we start with a state $f$ labeled by $(j,m_0)$  with $-j<
m_0<1+j$, we find that all states obtained by operating with $I_+$
will have norms of the same sign. These are related to all the
states $I_-^n f$ by a sign change in the norm. 
Consequently, the states of the non-unitary representation can
be divided into two families. In each family, the states have norms of the
same sign, while the two families are related by a change in
sign in the norm.

\vspace{1cm}
\begin{center}
{\large Appendix B: A Formula for the Transfer Matrix}
\end{center}
\vspace{7mm}
\setcounter{equation}{0}
\setcounter{section}{2}

When the potential is symmetric about the center of each period, it
is convenient to consider even and odd solutions $g(E,x)$, $u(E,x)$
such that
\begin{eqnarray}
g(E,0)=1,& g'(E,0)=0, \\
u(E,0)=0,& u'(E,0)=1.
\end{eqnarray}

Let us define \cite{james}

\begin{eqnarray}
g(E,-\frac{a}{2}) = g(E, \frac{a}{2}) = g_0 (E); \\
g'(E,-\frac{a}{2}) = -g'(E, \frac{a}{2}) = g'_0 (E); \\
u(E,-\frac{a}{2}) = -u(E, \frac{a}{2}) = u_0 (E); \\
u'(E,-\frac{a}{2}) = u'(E, \frac{a}{2}) = u'_0 (E);
\end{eqnarray}

According to the definition of transfer matrix \cite{cohen}, we can
derive a formula as follows:

\begin{equation}
T \, = \, \left(
\begin{array}{cc}
\alpha & \beta \\
\beta^{*} & \alpha^{*}
\end{array}
\right)
\end{equation}
where
\begin{eqnarray}
\alpha &=& e^{-ika} [(g_0 u'_0 + g'_0 u_0) + i(u'_0 g'_0 /k- u_0 g_0 k)] \\
\beta &=& -i e^{ika} (u'_0 g'_0 /k + u_0 g_0 k)
\end{eqnarray}
and $k=\sqrt E$, $a$ is the period.


\newpage

\begin{thebibliography}{99}
\bibitem{1}  For a survey, see for example: {\sl Dynamical Groups and
    Spectrum Generating Algebras}, Eds. A.Barut, A. Bohm and Y.Ne'eman (World
    Scientific, Singapore, 1987).

\bibitem{2}  F. Iachello,  {\it Chem. Phys. Lett.} {\bf 78} 581
            (1981); F. Iachello and R. D. Levine, {\it
            J. Chem. Phys.} {\bf 77} 3046 (1982);
            F. Iachello and R. Levine, {\it Algebraic Theory of Molecules} 
            (Oxford Press, Oxford, 1995); F. Iachello and
           A. Arima, {\sl The Interacting Boson Model}
           (Cambridge Press, Cambridge, 1987).
                                      

\bibitem{AGI}  Y. Alhassid, F. G\"{u}rsey and F. Iachello, {\it
    Phys. Rev. Lett.} {\bf     50} 873 (1983).

\bibitem{fi1} See for example, F. Iachello,  {\sl Rev. Nuovo
              Cimento} {\bf 19} 1 (1996), and references there in.

\bibitem{dk} D. Kusnezov, {\sl Phys. Rev. Lett.}  {\bf 79} 537 (1997).

\bibitem{li} H. Li and D. Kusnezov, Yale Univ. preprint (1999);
  {\it ibid}, in {\it Group 22: International Colloquium on Group
  Theoretical Methods in Physics}, Eds. S.P. Corney,
  R. Delbourgo and P.D. Jarvis, (International, Cambridge, MA,
  1999), p. 310.

\bibitem{scarf}  F. L. Scarf, {\it Phys. Rev.} {\bf  112} 1137 (1958).

\bibitem{fi2} Y. Alhassid, F. G\"ursey and F. Iachello,
  {\it Ann. Phys. (NY)} {\bf 167} 181 (1983); A. Frank and K. B. Wolf,
  {\it Phys. Rev. Lett.} {\bf 52} 1737 (1984).

\bibitem{gesztesy1}  F. Gesztesy and W. Kirsch,  {\it Journal 
            f\"{u}r Mathematik} {\bf 362} 28 (1984).


\bibitem{gursey}  F. G\"{u}rsey, in {\it Group Theoretical
                 Methods in Physics XI},
                 (Springer-Verlag, Berlin, 1983) p.106.

\bibitem{kittel} C. Kittel, {\it Quantum Theory of Solids},
                 (Wiley, New York, 1963).

\bibitem{bargmann}  V. Bargmann, {\it Ann. Math.} {\bf 48} 568 (1947).

\bibitem{pukanszky}  L. Puk\'{a}nszky, {\it Math. Annalen} {\bf
    156} 96 (1964); A. O. Barut and C. Fronsdal, {\it
    Proc. Roy. Soc. London} {\bf A287}, 532 (1965).

\bibitem{james}  H. M. James, {\it Phys. Rev.} {\bf 76} 1602 (1949).

\bibitem{cohen} C. Cohen-Tannoudji, B. Diu and F. Laloe,
      {\it Quantum Mechanics} (Wiley, New York, 1977) Vol.1 .

\bibitem{poschl} G. P\"oschl and E. Teller, {\it Z. Phys.} {\bf
    83} 143 (1933); L.D. Salem and R. Montemayor, {\it
    Phys. Rev.} {\bf A47} 105 (1993).
\bibitem{gesztesy2}  F. Gesztesy, C. Macdeo and L. Streit,  {\it
            J. Phys. A: Math. Gen.}{\bf 18} L503 (1985).

\bibitem{miller} W. Miller Jr., {\it Lie Theory and Special
    Functions}, (Academic Press, New York, 1968).

\end{thebibliography}
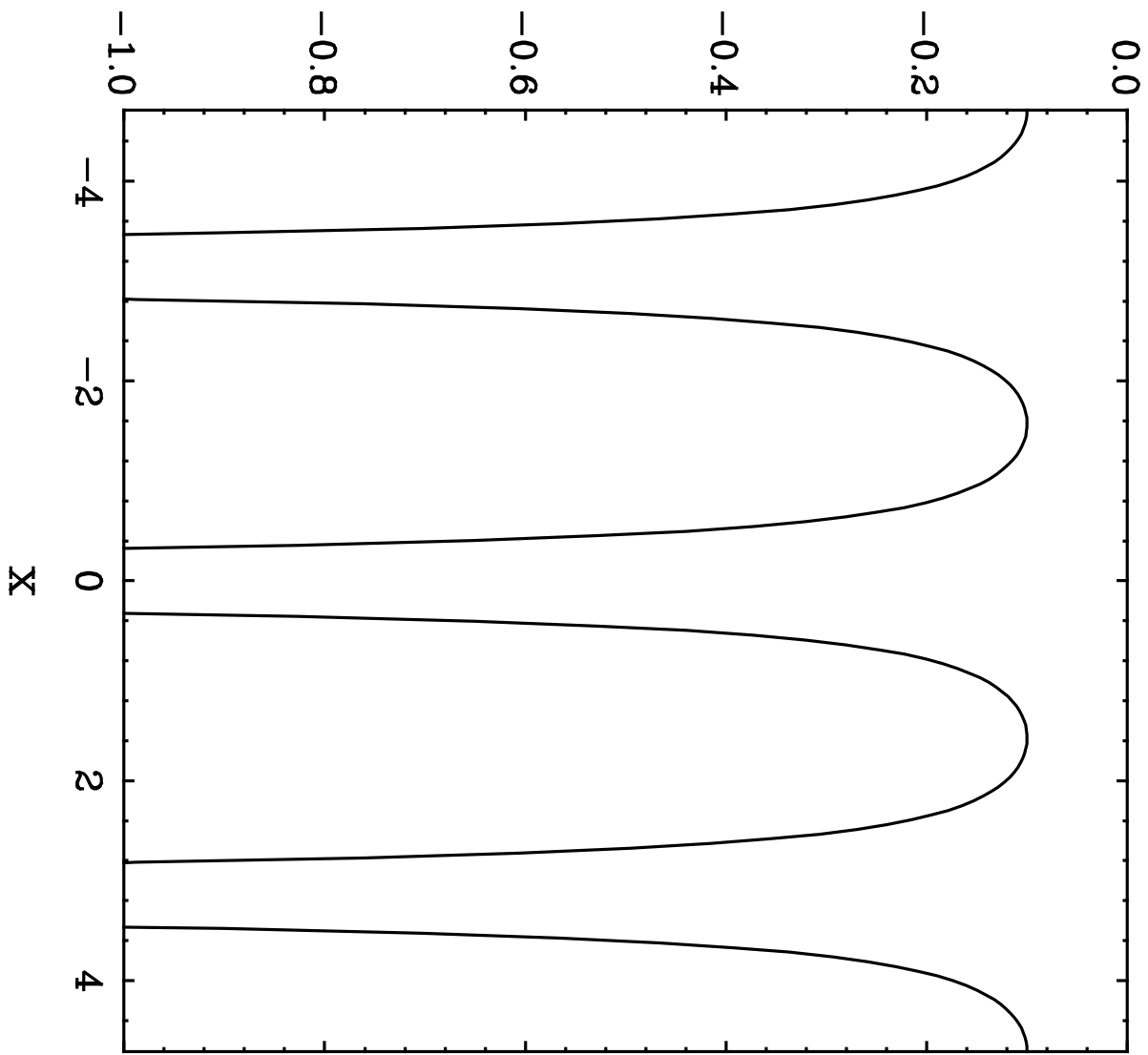
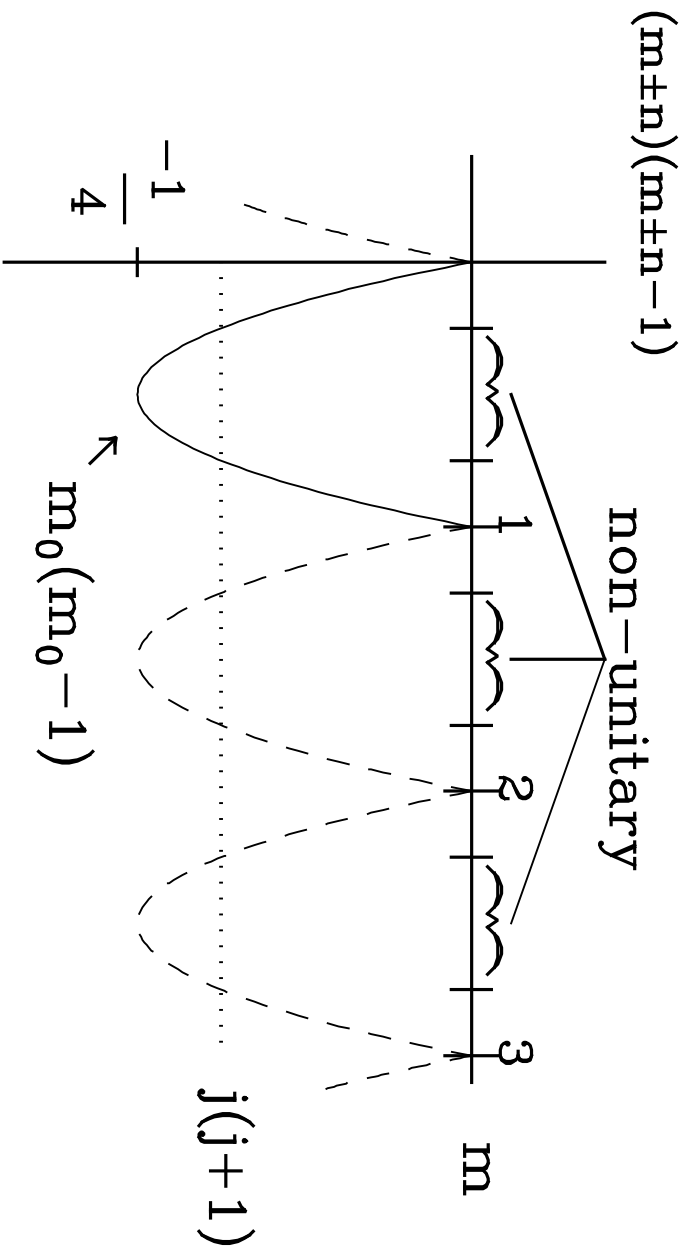
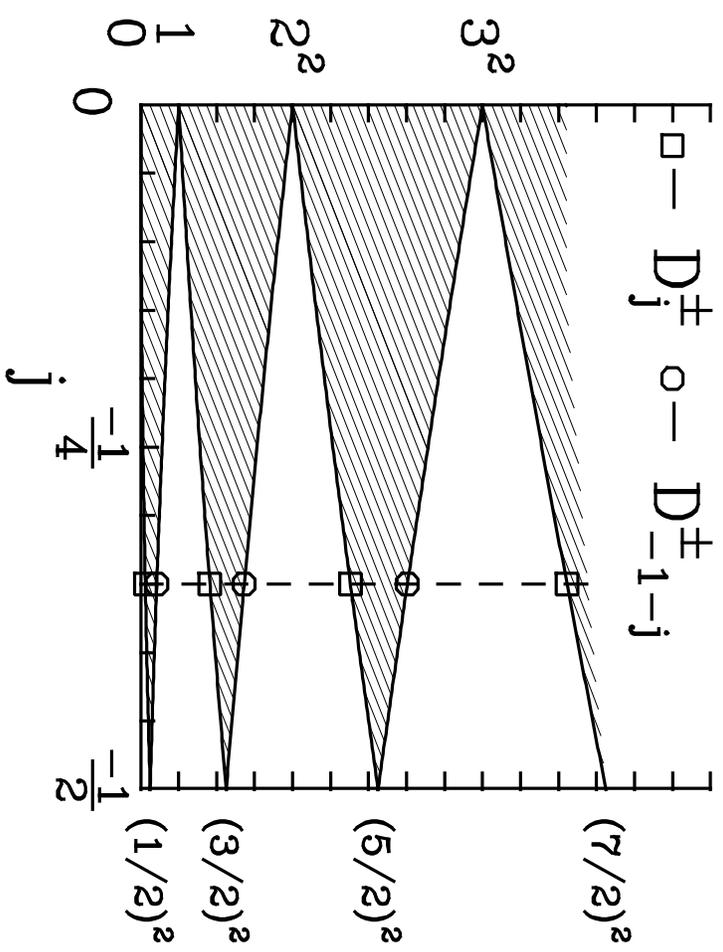
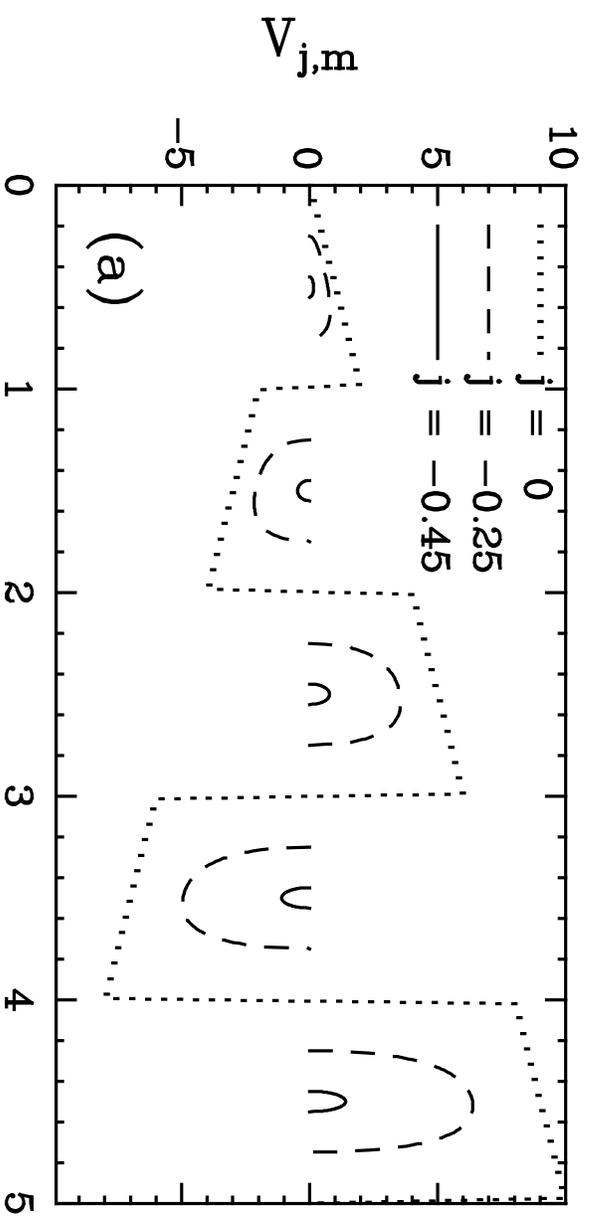
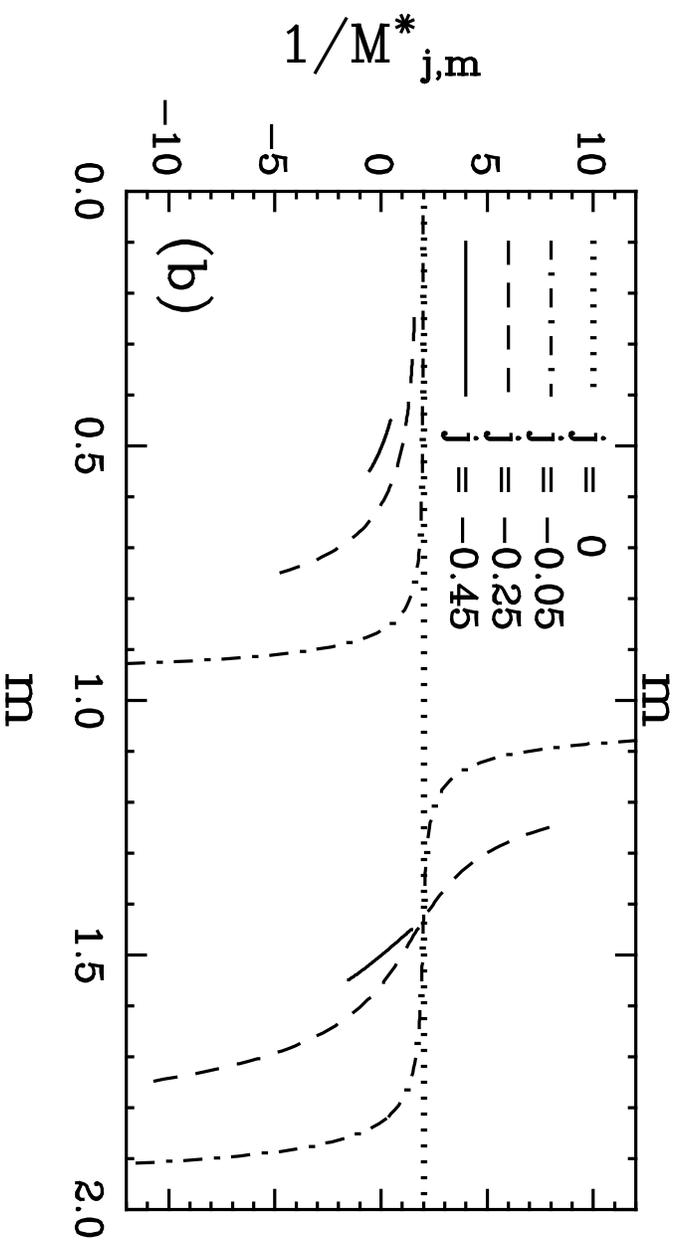
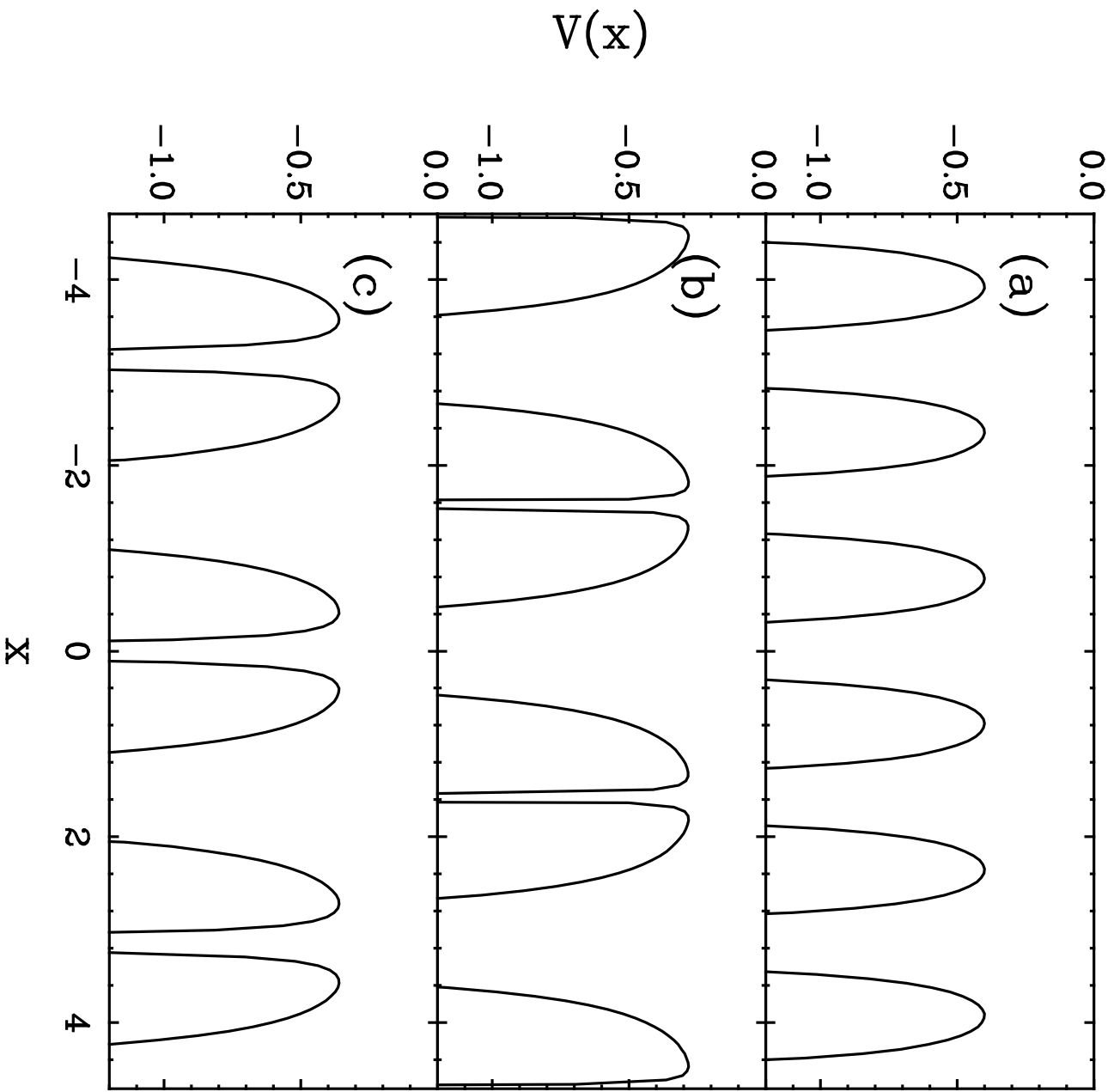
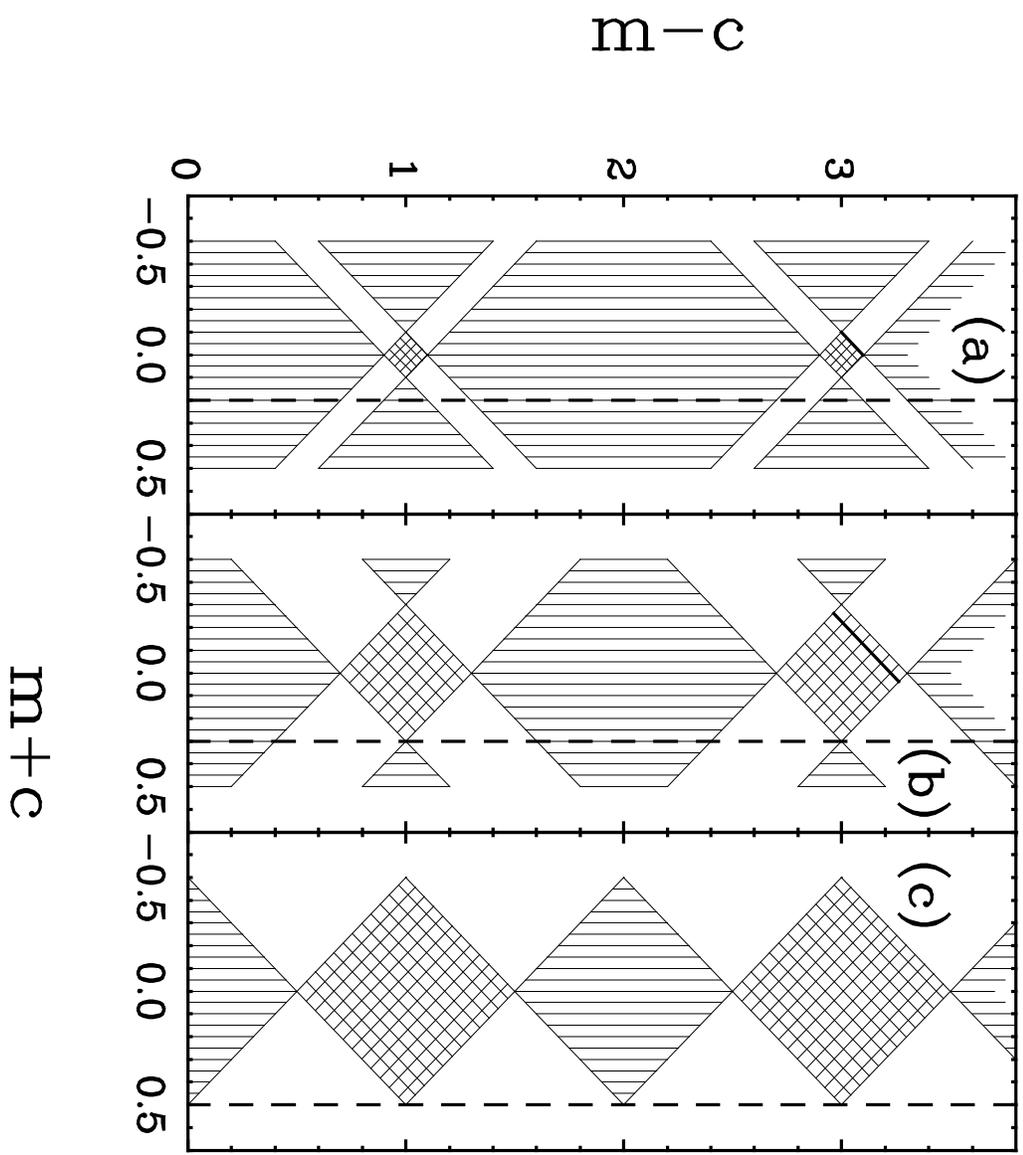
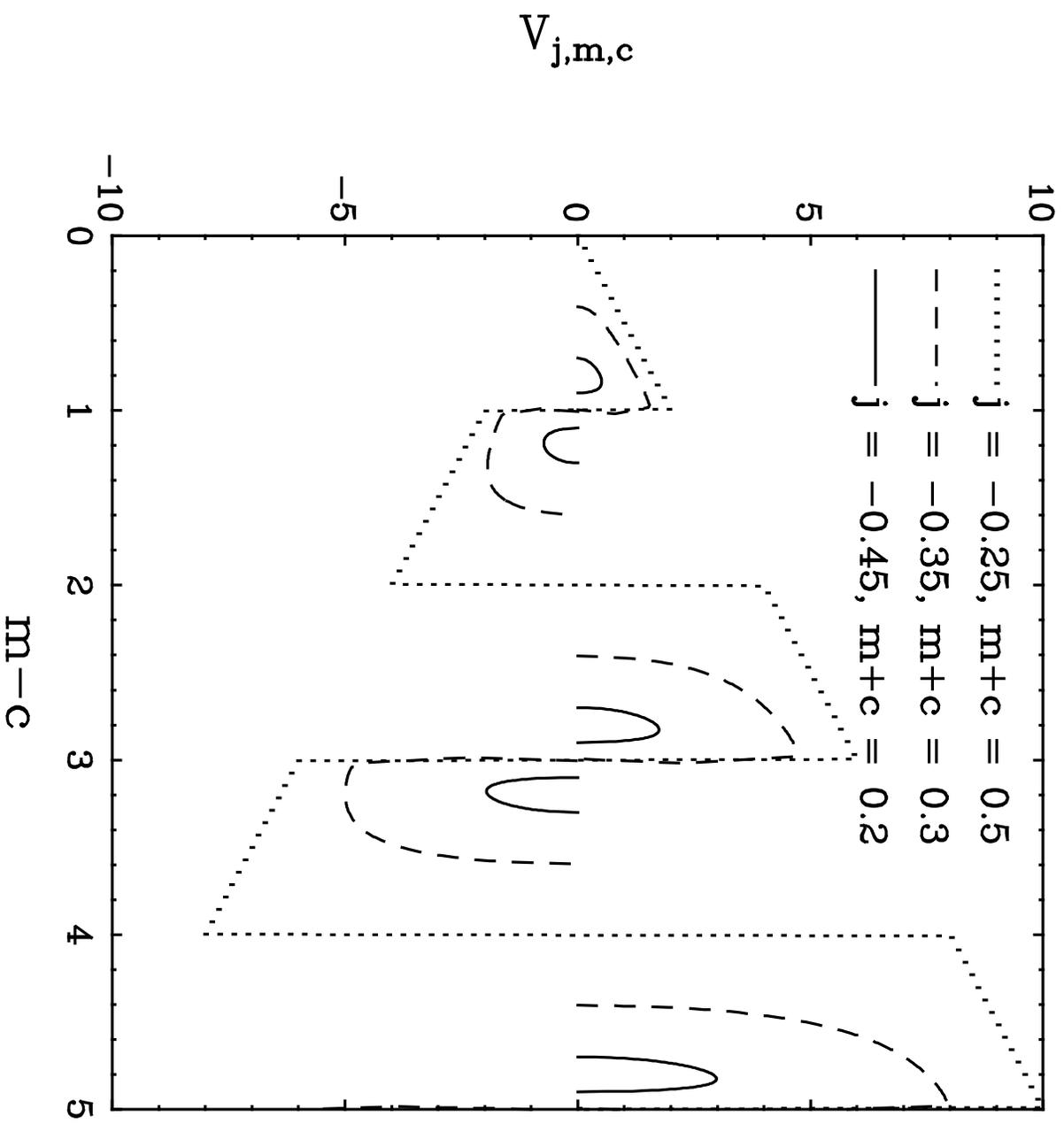
\end{document}